\documentclass[preprint,showpacs,preprintnumbers,amsmath,amssymb]{revtex4}

\usepackage{graphicx}

\begin{document}

\title{
Gravitational wave forms for a three-body system in Lagrange's orbit: 
parameter determinations and a binary source test  
}
\author{Hideki Asada} 
\email{asada@phys.hirosaki-u.ac.jp}
\affiliation{
Faculty of Science and Technology, Hirosaki University,
Hirosaki 036-8561, Japan} 

\date{\today}

\begin{abstract}
Continuing work initiated in an earlier publication 
[Torigoe et al. Phys. Rev. Lett. {\bf 102}, 251101 (2009)], 
gravitational wave forms for a three-body system 
in Lagrange's orbit are considered especially in an analytic method. 
First, we derive an expression of the three-body wave forms 
at the mass quadrupole, octupole and current quadrupole orders. 
By using the expressions, 
we solve a gravitational-wave {\it inverse} problem of 
determining the source parameters 
to this particular configuration (three masses, 
a distance of the source to an observer, and 
the orbital inclination angle to the line of sight) 
through observations of the gravitational wave forms alone. 
For this purpose, the chirp mass to a three-body system 
in the particular configuration is expressed in terms of 
only the mass ratios by deleting initial angle positions.  
We discuss also whether and how a binary source 
can be distinguished from a three-body system 
in Lagrange's orbit or others. 
\end{abstract}


\maketitle

\section{Introduction}
``{\it Can one hear the shape of a drum?}'' is a famous question 
posed by Mark Kac in 1966 \cite{Kac}, 
where to ``hear'' the shape of a drum is to infer information 
about the shape of the drumhead from the sound it makes. 
Such a question can be traced back to Hermann Weyl 
\cite{Weyl1911,Weyl1912}. 
Now, it is interesting to pose a gravitational-wave {\it inverse}  
problem for the forthcoming gravitational-wave astronomy 
by ground-based or space-borne detectors 
\cite{Centrella,Tinto,Mioa,Miob,LIGO-GRB,TAMA}. 
To ``hear'' a source through gravitational-wave observations 
is to extract the information 
about the source from the gravitational waves it makes. 

It is of general interest to ask 
``{\it can one tell how many apples are falling 
in the dark of night?}'' 
One simpler question is how and whether two-body 
and three-body gravitating systems can be distinguished 
through observations of gravitational waves that 
are made by these sources. 
Recently, this issue has been addressed \cite{THA}. 
They found that there is a case that 
quadrupole wave forms for two-body 
and three-body gravitating systems (in a particular configuration) 
cannot be distinguished even with observing the chirp, 
though different numbers of self-gravitating particles  
(in different types of periodic motion) are most likely 
to generate very different shapes of gravitational waves. 
In order to break this degeneracy between two-body and 
three-body sources in the particular configuration 
for the quadrupolar wave (even with observing frequency sweep), 
they suggested that higher multipolar wave forms 
(octupole for their cases) are needed. 

The purpose of the present paper is to investigate 
gravitational wave forms for a three-body system in 
the particular configuration (as Lagrange's orbit) 
in more detail, especially in analytic manners 
up to the mass octupole and current quadrupole orders. 
In particular, we shall discuss whether and how   
the source parameters (three masses, 
a distance of the sources to an observer, and 
their orbital inclination angle to the line of sight) 
can be determined 
through observations of the gravitational waveforms alone.

Inspiraling and finally merging binary compact stars are 
most likely astrophysical sources 
for the forthcoming direct detections of gravitational ripples 
(and consequently gravitational waves astronomy) 
by a lot of efforts by the on-going or designed detectors 
\cite{Centrella,Tinto,Mioa,Miob,LIGO-GRB,TAMA}. 
Merging neutron stars and black holes have been successfully 
simulated by numerical relativity 
\cite{Shibata,Pretorius,BBH06a,BBH06b,BBH06c}. 
Furthermore, analytic methods also have provided 
accurate waveform templates for 
inspiraling compact binaries, 
by post-Newtonian approaches 
(See \cite{Blanchet,FI} for reviews) 
and also by black hole perturbation techniques 
especially at the linear order in mass ratio  
(See also \cite{ST} for reviews). 
Bridges between the inspiraling stage and final merging phase 
are currently under construction (e.g., \cite{DN08a,DN08b}). 

There is a growing interest in potential astrophysical
sources of gravitational waves involving 3-body interactions 
(e.g., \cite{CIA,LN,ICA,THA} and references therein). 
Even the classical three-body (or N-body) 
problem in Newtonian gravity admits an infinite number of solutions, 
some of which express regular orbits and the others are chaotic, 
and in fact an increasing number of periodic orbits are found 
\cite{Danby,Marchal}. 
For the sake of simplicity, we focus on Lagrange's equilateral 
triangle solution, mostly because it offers a nice model tractable 
completely by hand (See below for more details).  

The linear perturbation analysis in Newton gravity \cite{Danby} 
shows that Lagrange triangular points $L_4$ and $L_5$ 
to the restricted three-body problem, 
in which one of three bodies is assumed as a test mass 
(e.g., an asteroid in the solar system), 
is stable if the mass ratio of the remaining two bodies 
is less than 0.0385, 
though relativistic corrections to stability of the system 
are poorly understood. 
Indeed, the ratio of the Jovian mass to the solar mass 
is $O(10^{-3})$. 

Here, it is worthwhile to mention previous works.  
Nakamura and Oohara \cite{NO} studied numerically the luminosity of 
gravitational radiation by N test particles 
orbiting around a Schwarzschild black hole, 
as an extension of Detweiler's analysis of the $N=1$ case 
\cite{Detweiler} 
by using Teukolsky equation \cite{Teukolsky}, 
in order to show the phase cancellation effect, 
which had been pointed out by Nakamura and Sasaki \cite{NS}. 
It should be noted that their N particles are test masses 
but not self-gravitating. 
The aim and setting of the present paper are different from 
those previous works. 

In this paper, we shall present an expression  
of gravitational waves for a three-body system 
in Lagrange's orbit,  
especially as a function of mass ratios. 
In addition, we shall discuss how to determine 
the source parameters (three masses, 
a distance of the source to an observer, and 
the orbital inclination angle to the line of sight) 
through observations of the gravitational waveforms alone.
We also discuss a possible test; which source is a binary, 
a three-body system in Lagrange's orbit or others?  

This paper is organized as follows. 
In section 2, we shall briefly summarize a notation and formulation 
for describing a three-body system in Lagrange's orbit 
and its gravitational waves. 
Wave forms at the mass quadrupole, mass octupole and 
current quadrupole orders are obtained in an analytic method. 
In section 3, we discuss a method of determining 
the source parameters from gravitational-wave observations alone. 
A binary source test is also discussed. 
Section 4 is devoted to the conclusion. 
In Appendix, we shall present calculations for 
the chirp mass and some useful relations. 
Throughout this paper, we take the units of $G=c=1$. 
Latin indices take 1, 2, 3, except for $p$ which labels each body. 

\section{Notation and Formulation}
\subsection{Linearized gravitational waves}
For a wave propagation direction denoted by a unit vector $n^a$, 
we define the transverse-traceless projection operator as 
\begin{equation}
P_{a}^{b}=\delta_{a}^{b}-n_an^b ,  
\label{P}
\end{equation}
where $\delta_{a}^{b}$ denotes the Kronecker's delta symbol. 
The linearized waves in the wave zone are  
expressed in terms of 
mass and current multipole moments denoted by 
$I^{A_{\ell}}$ and $S^{A_{\ell}}$, respectively \cite{Thorne}. 
These radiative multipole moments at the Newtonian order are 
related with the source position $x^a$, mass density $\rho$ 
and velocity $v^a$ as 
\begin{eqnarray}
I^{A_{\ell}} &=& \left[ \int \rho X^{A_{\ell}} d^3x \right]^{STF} , 
\label{I}
\\
S^{A_{\ell}} &=& \left[ \int \epsilon^{a_{\ell}}_{\; \; bc} 
x^b \rho v^c X^{A_{\ell-1}} d^3x \right]^{STF} , 
\label{S}
\end{eqnarray}
where $STF$ denotes the symmetric tracefree part, 
$\epsilon_{abc}$ denotes the Levi-Civita symbol 
in a three-dimensional Euclidean space, 
and 
we define the product of $\ell$ spatial coordinates as 
$X^{A_{\ell}} \equiv x^{a_1}x^{a_2} \cdots x^{a_{\ell}}$. 
We consider a system of spherical bodies 
approximated by massive particles. 
Then, Eqs. ($\ref{I}$) and ($\ref{S}$) become 
\begin{eqnarray}
I^{A_{\ell}} &=& \left[ \sum_{p=1}^N m_p X_{p}^{A_{\ell}} \right]^{STF} , 
\label{I2}
\\
S^{A_{\ell}} &=& \left[ \sum_{p=1}^N m_p \epsilon^{a_{\ell}}_{\; \; bc} 
x_{p}^{b} v_{p}^{c} X_{p}^{A_{\ell-1}} \right]^{STF} , 
\label{S2}
\end{eqnarray}
where $N$ denotes the number of the particles 
and the subscript $p$ denotes the $p$-th body. 

In terms of the multipole moments, 
the linearized waves at the wave zone 
are expressed as \cite{Thorne} 
\begin{eqnarray}
h_{jk}^{TT}(t, \mbox{\boldmath $x$})
&=&\frac{1}{r}
\left[ 
\sum_{\ell=2}^{\infty} \left(\frac{4}{\ell !} 
I_{jkA_{\ell-2}}^{(\ell)}\right) (t-r) N^{A_{\ell-2}} \right. 
\nonumber\\
& & \left. + \sum_{\ell=2}^{\infty} \left(\frac{8\ell}{(\ell+1) !} 
\epsilon^b_{\; c(j} S_{k)bA_{\ell-2}}^{(\ell) }\right) (t-r) n^c N^{A_{\ell-2}} 
\right]^{TT} +O\left(\frac{1}{r^2}\right) , 
\label{hTT}
\end{eqnarray}
where $t$ and $r$ mean time and source distance,
respectively, in the Minkowskian spherical coordinates 
$(t, r, \Theta, \Phi)$, 
$TT$ denotes the transverse-traceless part, 
$(\ell)$ denotes the $\ell$-th time derivative, 
and 
we define the tensor product of $\ell-2$ unit radial vectors as 
$N^{A_{\ell-2}} \equiv n^{a_1}n^{a_2} \cdots n^{a_{\ell-2}}$. 

For a binary case, mass quadrupolar, octupolar and 
current quadrupolar waves were considered fully 
by Blanchet and Schafer \cite{BS}, 
where they showed that the octupolar part is linearly 
proportional to a mass difference.

\subsection{Lagrange's solution} 
Let us consider the Lagrange's solution for a three-body system   
(on $x$-$y$ plane), where each mass is denoted 
by $m_p$ $(p=1, 2, 3)$.   
The initial positions of each mass are expressed as 
$\mbox{\boldmath $x$}_1=(0, 0)$, 
$\mbox{\boldmath $x$}_2=a (\sqrt{3}/2, 1/2)$, 
and 
$\mbox{\boldmath $x$}_3=a (0, 1)$, 
where the side of a regular triangle is denoted as $a$ 
\cite{Danby,THA}. 

We choose the spatial coordinates such that 
the center of mass (COM) is at rest as 
$(x_{\rm COM}, y_{\rm COM})
= a (\sqrt{3} {\nu}_2/2, ({\nu}_2+{\nu}_3)/2)$), 
where the total mass and mass ratio are denoted 
as $m_{\rm tot} \equiv \sum_p m_p$ and 
${\nu}_p \equiv m_p/m_{\rm tot}$, respectively. 
We have an identity as 
\begin{equation}
\nu_1 + \nu_2 + \nu_3 = 1 . 
\label{unity}
\end{equation}
The orbital frequency $\omega$ for the triangle 
satisfies 
\begin{equation}
\omega^2 = \frac{m_{\rm tot}}{a^3} , 
\label{Kepler}
\end{equation} 
which takes the same form as Kepler's third law for a binary system 
but with the total mass of three masses. 

\begin{figure}[t]
\includegraphics[width=10cm]{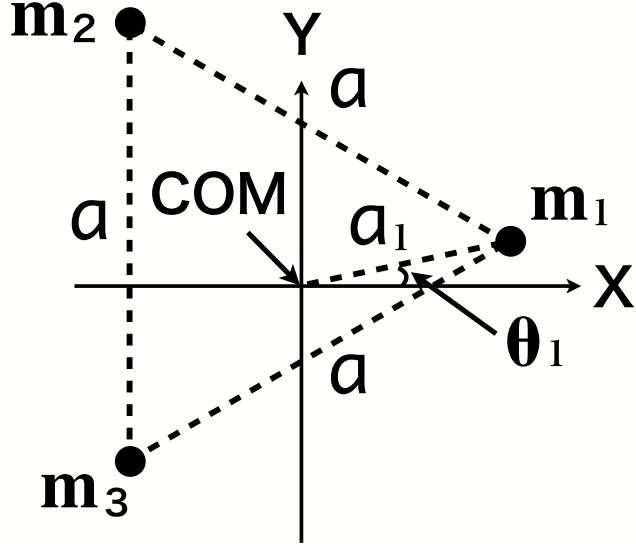}
\caption{
Definition of $\theta_p$ 
in the Lagrange's orbit. 
The angle $\theta_p$ is measured from $X$-axis 
to the direction of each mass at the initial time.  
}
\label{f1}
\end{figure}

Henceforth, it is convenient to employ the COM coordinates 
$(X, Y)$ that can be obtained by a translation from $(x, y)$. 
In the COM coordinates, the location of each mass  
is expressed as 
\begin{equation}
\mbox{\boldmath $X$}_p = 
a_p (\cos(\omega t+\theta_p), \sin(\omega t+\theta_p)), 
\label{Xp}
\end{equation}
where $a_p$ is defined as 
$a_1=\sqrt{x_{\rm COM}^2+y_{\rm COM}^2}$, 
$a_2=\sqrt{(3^{1/2}a/2-x_{\rm COM})^2+(a/2-y_{\rm COM})^2}$, 
and 
$a_3=\sqrt{x_{\rm COM}^2+(a-y_{\rm COM})^2}$, respectively, 
and $\theta_p$ denotes the angle between the new $X$-axis 
and the direction of each mass at $t=0$ 
(See Fig. $\ref{f1}$). 
In practice, computations can be simplified by using 
complex variables. 
In particular, variables for the triangle configuration 
$\theta_p$ are written in terms of the mass ratios 
(See Appendix for more detail). 

\section{Wave forms for three bodies in Lagrange's orbit}
Let $i$ denote the orbital inclination angle  
with respect to the line of sight.  
\subsection{Mass Quadrupole}
By direct calculations, we obtain the plus mode of 
quadrupole waves as 
\begin{eqnarray}
r\times h_{\rm Q}^{+} 
&=& 
-2 \sum_p m_p a_p^2 \omega^2 (1+\cos^2 i) 
\cos 2(\omega t+\theta_p) ,  
\label{hQplus0}
\end{eqnarray}
where the subscript $Q$ denotes a mass quadrupolar part. 
By using Eqs. ($\ref{a1}$)-($\ref{a3}$) and 
($\ref{e2itheta1}$)-($\ref{e2itheta3}$), 
Eq. ($\ref{hQplus0}$) is rewritten as 
\begin{eqnarray}
r\times h_{\rm Q}^{+} 
&=& 
-m_{\rm tot} a^2 \omega^2 (1+\cos^2 i) 
\nonumber\\
&&
\times 
\left[
\Bigl(\nu_1 (\nu_2+\nu_3) - 2 \nu_2 \nu_3 \Bigr) \cos 2\omega t 
+ 
\sqrt{3} \nu_1 (\nu_2-\nu_3) \sin 2\omega t 
\right] . 
\label{hQplus}
\end{eqnarray}
In a similar manner, we obtain the cross-mode as 
\begin{eqnarray}
r\times h_{\rm Q}^{\times} 
&=& 
-4 \sum_p m_p a_p^2 \omega^2 \cos i  
\sin 2(\omega t+\theta_p)  
\nonumber\\
&=& 
-2 m_{\rm tot} a^2 \omega^2 \cos i 
\nonumber\\
&&
\times 
\left[
\Bigl( \nu_1 (\nu_2+\nu_3) - 2 \nu_2 \nu_3 \Bigr) \sin 2\omega t 
- 
\sqrt{3} \nu_1 (\nu_2-\nu_3) \cos 2\omega t 
\right] .  
\label{hQcross}
\end{eqnarray}

For a three-body system in Lagrange's orbit, 
the frequency sweep due to gravitational radiation reaction 
has been recently obtained as \cite{THA} 
\begin{eqnarray}
\frac{1}{f_{\rm GW}}\frac{df_{\rm GW}}{dt}
&=&
\frac{96}{5} \pi^{8/3} M_{\rm chirp}^{5/3} f_{\rm GW}^{8/3} ,  
\label{df}
\end{eqnarray}
where we define a chirp mass as 
\begin{eqnarray}
&&M_{\rm chirp} 
\nonumber\\
&&= 
m_{\rm tot}  
\left[ 
\frac{
\left\{
\sum_p \nu_p \left(\frac{M^{\rm eff}_p}{m_{\rm tot}}\right)^{2/3} 
\right\}^2 
-2 \sum_{p \neq q} \nu_p \nu_q 
\left(\frac{M^{\rm eff}_p}{m_{\rm tot}}\right)^{2/3} 
\left(\frac{M^{\rm eff}_q}{m_{\rm tot}}\right)^{2/3} 
\sin^2(\theta_p - \theta_q) 
}
{
\sum_{p \neq q} \nu_p \nu_q 
- \sum_p \nu_p \left(\frac{M^{\rm eff}_p}{m_{\rm tot}}\right)^{2/3} 
}
\right]^{3/5} , 
\label{Mchirp-old}
\end{eqnarray} 
and $M^{\rm eff}_p$ denotes the {\it effective} one-body mass 
for which the equation of motion becomes \cite{Danby} 
\begin{equation}
\frac{d^2 \mbox{\boldmath $X$}_p}{dt^2}  
= 
-\frac{M^{\rm eff}_p \mbox{\boldmath $X$}_p}
{|\mbox{\boldmath $X$}_p|^3} . 
\label{EOM}
\end{equation}
For instance, $M^{\rm eff}_1$ is defined as 
\begin{equation}
M^{\rm eff}_1
= 
\frac{(m_2^2+m_2m_3+m_3^2)^{3/2}}{m_{\rm tot}^2} . 
\label{Meff}
\end{equation}
By cyclic permutations, $M^{\rm eff}_2$ and $M^{\rm eff}_3$ are defined. 
It is worthwhile to mention that 
this frequency evolution equation is the same as 
that for a binary system \cite{THA} including the numerical
coefficient $96/5$. 

The chirp mass is expressed in terms of 
both each mass $m_p$ and its initial position angle $\theta_p$. 
It is rewritten as a function only of the mass ratios 
by deleting the initial position angles as 
(See Appendix for detailed calculations) 
\begin{equation}
M_{\rm chirp} = m_{\rm tot} \times F(\nu_1, \nu_2, \nu_3) , 
\label{mchirp}
\end{equation}
where we define $F$ as 
\begin{eqnarray}
F&=&
\left(
\frac12
\frac{\nu_1^2 (\nu_2-\nu_3)^2 + \nu_2^2 (\nu_3-\nu_1)^2 
+ \nu_3^2 (\nu_1-\nu_2)^2}{\nu_1\nu_2+\nu_2\nu_3+\nu_3\nu_1}
\right)^{3/5} . 
\label{F}
\end{eqnarray}

By using this expression of the three-body chirp mass 
for Eqs. ($\ref{hQplus}$) and ($\ref{hQcross}$), 
we obtain 
\begin{eqnarray}
r\times h_{\rm Q}^{+} 
&=&
-2 m_{\rm tot}^{5/6}  M_{\rm chirp}^{5/6} \omega^{2/3} (1+\cos^2 i) 
\nonumber\\
&&
\times 
\sqrt{\nu_1 \nu_2 + \nu_2 \nu_3 + \nu_3 \nu_1} 
\cos (2\omega t - \Psi_{\rm Q}) , 
\label{hQplusPsi}
\end{eqnarray}
where we define $\Psi_{\rm Q}$ as 
\begin{equation}
\tan\Psi_{\rm Q} = \frac{B_{\rm Q}}{A_{\rm Q}} , 
\label{PsiQ}
\end{equation}
for 
\begin{eqnarray}
A_{\rm Q}&=&\nu_1 (\nu_2+\nu_3) - 2 \nu_2 \nu_3 , 
\label{AQ}\\
B_{\rm Q}&=&\sqrt{3} \nu_1 (\nu_2-\nu_3) . 
\label{BQ}
\end{eqnarray}
We rewrite the cross mode as 
\begin{eqnarray}
r\times h_{\rm Q}^{\times} 
&=&
-2 m_{\rm tot}^{5/6}  M_{\rm chirp}^{5/6} \omega^{2/3} \cos i 
\nonumber\\
&&
\times 
\sqrt{\nu_1 \nu_2 + \nu_2 \nu_3 + \nu_3 \nu_1} 
\sin (2\omega t - \Psi_{\rm Q}) .  
\label{hQcrossPsi}
\end{eqnarray}

\subsection{Mass Octupole}
We obtain, by direct calculations, the plus mode of 
quadrupole waves as 
\begin{eqnarray}
r\times h_{\rm Oct}^{+} 
&=& 
-\frac{1}{12} \sum_p m_p a_p^3 \omega^3 \sin i 
\nonumber\\
&&
\times 
\left[
27 (1+\cos^2i) \cos 3(\omega t+\theta_p) 
+ (1-3\cos^2i) \cos (\omega t+\theta_p) 
\right] ,  
\label{hOctplus0}
\end{eqnarray}
where the subscript $Oct$ denotes 
a mass octupolar part.  

By using Eqs. ($\ref{a1}$)-($\ref{a3}$),  
($\ref{eitheta1}$)-($\ref{eitheta3}$) and 
($\ref{e3itheta1}$)-($\ref{e3itheta3}$), 
one can rewrite Eq. ($\ref{hOctplus0}$) as 
\begin{eqnarray}
r\times h_{\rm Oct}^{+} 
&=& 
-\frac{1}{12} m_{\rm tot}^2 \omega \sin i  
\nonumber\\
&&
\times 
\Bigl[
27 (1+\cos^2 i) 
\left( 3^{3/2} \nu_1 \nu_2 \nu_3 \cos 3\omega t 
+ (\nu_1 - \nu_2) (\nu_2 - \nu_3) (\nu_3 - \nu_1) 
\sin 3 \omega t 
\right) 
\nonumber\\
&&
+ 
(1-3\cos^2 i) 
\left( 
\frac{\sqrt3}{2} \nu_1 
\{ \nu_2 (\nu_2 - \nu_1) + \nu_3 (\nu_3 - \nu_1) \} 
\cos \omega t 
\right. 
\nonumber\\
&&
\left.
-\frac12 (\nu_2 - \nu_3) 
\{ (\nu_1 - \nu_2) (\nu_1 - \nu_3) - 3 \nu_2 \nu_3 \} 
\sin \omega t 
\right) 
\Bigr] , 
\label{hOctplus}
\end{eqnarray}
where we use Eq. ($\ref{Kepler}$).  

We obtain the cross mode of mass octupole waves as 
\begin{eqnarray}
r\times h_{\rm Oct}^{\times} 
&=& 
-\frac{1}{12} \sum_p m_p a_p^3 \omega^3 \sin 2 i  
\nonumber\\
&&
\times 
\left[
27 \sin 3(\omega t+\theta_p) 
- \sin (\omega t+\theta_p) 
\right] . 
\label{hOctcross0}
\end{eqnarray}
By using Eqs. ($\ref{a1}$)-($\ref{a3}$),  
($\ref{eitheta1}$)-($\ref{eitheta3}$) and 
($\ref{e3itheta1}$)-($\ref{e3itheta3}$), 
one can rewrite Eq. ($\ref{hOctcross0}$) as 
\begin{eqnarray}
r\times h_{\rm Oct}^{\times} 
&=& 
-\frac{1}{12} m_{\rm tot}^2 \omega \sin 2 i  
\nonumber\\
&&
\times 
\Bigl[
27 
\Bigl( 3^{3/2} \nu_1 \nu_2 \nu_3 \sin 3\omega t 
- (\nu_1 - \nu_2) (\nu_2 - \nu_3) (\nu_3 - \nu_1) 
\cos 3 \omega t 
\Bigr) 
\nonumber\\
&&
- \left(
\frac{\sqrt3}{2} \nu_1 
\{ \nu_2 (\nu_2 - \nu_1) + \nu_3 (\nu_3 - \nu_1) \} 
\sin \omega t 
\right. 
\nonumber\\
&&
\left.
+\frac12 (\nu_2 - \nu_3) 
\{ (\nu_1 - \nu_2) (\nu_1 - \nu_3) - 3 \nu_2 \nu_3 \} 
\cos \omega t 
\right) 
\Bigr] ,  
\label{hOctcross}
\end{eqnarray}
where we use Eq. ($\ref{Kepler}$).

\subsection{Current Quadrupole}
Here, we consider current quadrupolar waves. 
The plus mode becomes 
\begin{eqnarray}
r\times h_{\rm C}^{+} 
&=& 
\frac{4}{3} m_{\rm tot}^2 \omega \sin i  
\nonumber\\
&&
\times 
\Bigl[ 
\frac{\sqrt3}{2} \nu_1 
\{ \nu_2 (\nu_2 - \nu_1) + \nu_3 (\nu_3 - \nu_1) \} 
\cos \omega t 
\nonumber\\
&&
-\frac12 (\nu_2 - \nu_3) 
\{ (\nu_1 - \nu_2) (\nu_1 - \nu_3) - 3 \nu_2 \nu_3 \} 
\sin \omega t 
\Bigr] ,  
\label{hCplus}
\end{eqnarray}
where the subscript $C$ denotes a current quadrupolar part. 

The cross mode becomes 
\begin{eqnarray}
r\times h_{\rm C}^{\times} 
&=& 
\frac{2}{3} m_{\rm tot}^2 \omega \sin 2 i  
\nonumber\\
&&
\times 
\Bigl[ 
\frac{\sqrt3}{2} \nu_1 
\{ \nu_2 (\nu_2 - \nu_1) + \nu_3 (\nu_3 - \nu_1) \} 
\sin \omega t 
\nonumber\\
&&
+\frac12 (\nu_2 - \nu_3) 
\{ (\nu_1 - \nu_2) (\nu_1 - \nu_3) - 3 \nu_2 \nu_3 \} 
\cos \omega t 
\Bigr] . 
\label{hCcross}
\end{eqnarray}

Both the mass octupolar and current quadrupolar parts 
are proportional to $m_{\rm tot}^2 \omega$. 
Hence they can be combined as 
\begin{eqnarray}
r\times h_{\rm Oct+C}^{+} 
&=& 
-\frac{1}{4} m_{\rm tot}^2 \omega \sin i  
\nonumber\\
&&
\times 
\Bigl[
9 (1+\cos^2 i) 
\left( 3^{3/2} \nu_1 \nu_2 \nu_3 \cos 3\omega t 
+ (\nu_1 - \nu_2) (\nu_2 - \nu_3) (\nu_3 - \nu_1) 
\sin 3 \omega t 
\right) 
\nonumber\\
&&
- 
(5+\cos^2 i) 
\left( 
\frac{\sqrt3}{2} \nu_1 
\{ \nu_2 (\nu_2 - \nu_1) + \nu_3 (\nu_3 - \nu_1) \} 
\cos \omega t 
\right. 
\nonumber\\
&&
\left.
-\frac12 (\nu_2 - \nu_3) 
\{ (\nu_1 - \nu_2) (\nu_1 - \nu_3) - 3 \nu_2 \nu_3 \} 
\sin \omega t 
\right) 
\Bigr] , 
\label{hOct+Cplus}
\end{eqnarray}
and 
\begin{eqnarray}
r\times h_{\rm Oct+C}^{\times} 
&=& 
-\frac{1}{4} m_{\rm tot}^2 \omega \sin 2 i  
\nonumber\\
&&
\times 
\Bigl[
9 
\Bigl( 3^{3/2} \nu_1 \nu_2 \nu_3 \sin 3\omega t 
- (\nu_1 - \nu_2) (\nu_2 - \nu_3) (\nu_3 - \nu_1) 
\cos 3 \omega t 
\Bigr) 
\nonumber\\
&&
-3 \left(
\frac{\sqrt3}{2} \nu_1 
\{ \nu_2 (\nu_2 - \nu_1) + \nu_3 (\nu_3 - \nu_1) \} 
\sin \omega t 
\right. 
\nonumber\\
&&
\left.
+\frac12 (\nu_2 - \nu_3) 
\{ (\nu_1 - \nu_2) (\nu_1 - \nu_3) - 3 \nu_2 \nu_3 \} 
\cos \omega t 
\right) 
\Bigr] . 
\label{hOct+Ccross}
\end{eqnarray}

It is natural that one can recover the wave forms to 
a binary system from 
Eqs. ($\ref{hOct+Cplus}$) and ($\ref{hOct+Ccross}$) 
when a third mass vanishes, 
say $\nu_3=0$. 

\begin{figure}[t]
\includegraphics[width=15cm]{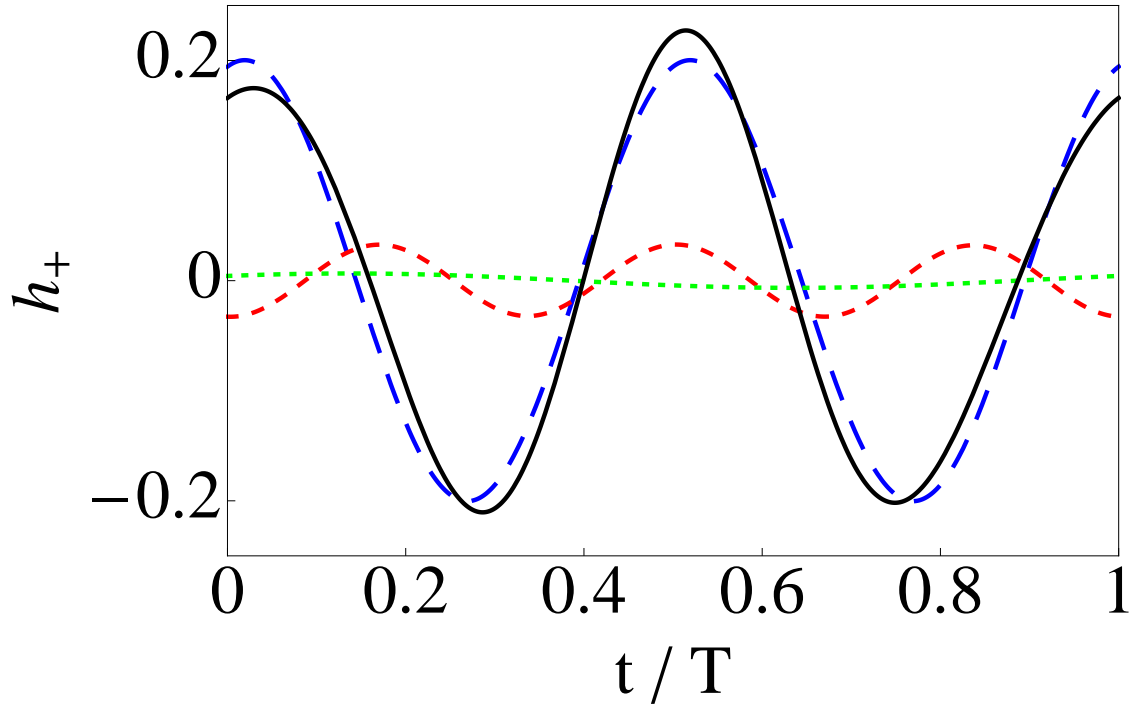}
\caption{
Gravitational waves for a three-body system in Lagrange's orbit. 
Log-dashed (blue), short-dashed (red) and dotted (green) curves 
denote mass quadrupolar, octupolar and current quadrupolar parts, 
respectively. 
The sold (black) one denotes the total wave forms. 
The vertical axis is in arbitrary units and 
time in the horizontal axis is normalized by the orbital period. 
For simplicity, the mass ratio is assumed as 
$m_1: m_2: m_3 = 1: 2: 3$, though stability arguments 
prefer much larger ratios \cite{Danby}. 
In order to exaggerate differences between each component of waves, 
we assume a mildly relativistic case of $a = 100m_{\rm tot}$, 
which corresponds to $v/c \sim 0.1$. 
The mass octupolar part makes a relatively large contribution 
mostly because of including a large numerical coefficient 
(as 27 in Eq. ($\ref{hOctplus}$)), 
whereas the current quadrupolar part is much smaller 
and changes slowly with time because it has no $3\omega$ 
but only the $\omega$ part. 
}
\label{f2}
\end{figure}

Figure $\ref{f2}$ shows wave forms due to mass quadrupole, 
octupole and current quadrupole parts that are 
expressed above. 

\section{Parameter determinations}
The frequency $\omega$ can be directly determined 
as $\omega = \pi f_{\rm GW}$ from measurements of the frequency 
of the mass quadrupolar wave $f_{\rm GW}$. 
What we have to do is to determine the remaining six quantities 
as $m_{\rm tot}$, $\nu_1$, $\nu_2$, $\nu_3$, $r$ and $i$. 
We have an identity $\nu_1 + \nu_2 + \nu_3 =1$. 
Hence, we need to find out five more relations 
for parameter determinations. 

The frequency of mass quadrupolar waves is $2\omega$, 
whereas that for the combination of mass octupolar 
and current quadrupolar ones is either $\omega$ or 
$3\omega$. 
By using this difference of frequency dependences, 
therefore, one can pick up the quadrupolar waves from 
observed signals. 
By comparing amplitudes of the $+$ and $\times$ modes, 
we obtain
\begin{equation}
\frac{Amp\left( h_{\rm Q}^{\times} \right)}
{Amp\left( h_{\rm Q}^{+} \right)} 
= \frac{2 \cos i}{1+\cos^2 i} ,  
\label{Amp-ratio}
\end{equation}
where $Amp$ denotes the amplitude of waves. 
This relation is the same as the well-known one for binaries. 
The L.H.S. of Eq. ($\ref{Amp-ratio}$) can be determined 
by observations and thus the R.H.S. tells us 
the inclination angle $i$.

Through observing the frequency sweep 
of the mass quadrupolar part (See Eq. ($\ref{df}$)), 
one can determine the chirp mass as 
\begin{equation}
M_{\rm chirp} 
= 
\left(
\frac{5}{96}
\frac{1}{\pi^{8/3}} 
\frac{1}{f_{\rm GW}^{11/3}}
\frac{d f_{\rm GW}}{d t}  
\right)^{3/5} ,  
\label{det-chirpmass}
\end{equation}
where $f_{\rm GW}=2 f$ for $f=\omega/2\pi$. 

Next, let the amplitude of each wave component 
observed separately. 
{}From Eq. ($\ref{hQplusPsi}$), 
first, we obtain that for the mass quadrupolar part as 
\begin{eqnarray}
r\times Amp\left( h_{\rm Q}^{+} \right)
&=&
2 m_{\rm tot}^{5/6}  M_{\rm chirp}^{5/6} \omega^{2/3} (1+\cos^2 i) 
\nonumber\\
&&
\times 
\sqrt{\nu_1 \nu_2 + \nu_2 \nu_3 + \nu_3 \nu_1} , 
\label{amp-hQ}
\end{eqnarray}
which can be solved for the distance $r$ as  
\begin{eqnarray}
r&=&\frac{2 m_{\rm tot}^{5/6} M_{\rm chirp}^{5/6} \omega^{2/3} (1+\cos^2 i) 
\sqrt{\nu_1\nu_2+\nu_2\nu_3+\nu_3\nu_1} }
{Amp\left( h_{\rm Q}^{+} \right)} . 
\label{det-r}
\end{eqnarray}
For parameter determinations, however, it is useful to 
rewrite this as 
\begin{eqnarray}
\frac{m_{\rm tot}^{5/6} \sqrt{\nu_1\nu_2+\nu_2\nu_3+\nu_3\nu_1}}{r} 
&=& 
\frac{Amp\left( h_{\rm Q}^{+} \right)}
{2 M_{\rm chirp}^{5/6} \omega^{2/3} (1+\cos^2 i)} , 
\label{det-r2}
\end{eqnarray}
where the R.H.S. is known up to this point. 

By using Eq. ($\ref{hOct+Cplus}$), 
the magnitude of $h_{\rm Oct+C}^{+}$ at $3\omega$ becomes  
\begin{eqnarray}
r\times Amp\left( h_{\rm Oct+C}^{+}|_{3\omega} \right) 
&=& 
\frac{9}{4} m_{\rm tot}^2 \omega \sin i (1+\cos^2 i)
\nonumber\\
&&
\times 
\sqrt{27 \nu_1^2\nu_2^2\nu_3^2 
+ (\nu_1-\nu_2)^2 (\nu_2-\nu_3)^2 (\nu_3-\nu_1)^2} , 
\label{amp-hOctC+3omega} 
\end{eqnarray}
where the subscript of $3\omega$ in the L.H.S. 
means a $3\omega$ part. 
This equation is rewritten as 
\begin{eqnarray}
&&\frac
{m_{\rm tot}^2 \sqrt{27 \nu_1^2\nu_2^2\nu_3^2 
+ (\nu_1-\nu_2)^2 (\nu_2-\nu_3)^2 (\nu_3-\nu_1)^2} }
{r} 
\nonumber\\
&& 
=
\frac{4}{9} 
\frac{1}{\omega \sin i (1+\cos^2 i)} 
Amp\left( h_{\rm Oct+C}^{+}|_{3\omega} \right) ,  
\label{det-OctC3omega} 
\end{eqnarray}
where the R.H.S. is known up to this point. 

By using Eq. ($\ref{hOct+Cplus}$), 
the magnitude of $h_{\rm Oct+C}^{+}$ at $\omega$ becomes  
\begin{eqnarray}
r\times Amp\left( h_{\rm Oct+C}^{+}|_{\omega} \right) 
&=& 
\frac{1}{8} m_{\rm tot}^2 \omega \sin i (5+\cos^2 i) 
\nonumber\\
&&
\times 
\left[
3 
\{ \nu_1 \nu_2 (\nu_1 - \nu_2) 
+\nu_2 \nu_3 (\nu_2 - \nu_3) 
+\nu_3 \nu_1 (\nu_3 - \nu_1) 
\}^2 
\right.
\nonumber\\
&&
\left.
\: \: \:
+
(\nu_1-\nu_2)^2 (\nu_2-\nu_3)^2 (\nu_3-\nu_1)^2
\right]^{1/2} , 
\label{amp-hOctC+omega} 
\end{eqnarray}
where the subscript of $\omega$ in the L.H.S. 
means a $\omega$ part. 
This equation is rewritten as 
\begin{eqnarray}
&&\frac{m_{\rm tot}^2}{r} 
\times 
\Bigl[
3 
\Bigl( \nu_1 \nu_2 (\nu_1 - \nu_2) 
+\nu_2 \nu_3 (\nu_2 - \nu_3) 
+\nu_3 \nu_1 (\nu_3 - \nu_1) 
\Bigr)^2 
\nonumber\\
&&
{}\;\;\;
+ 
(\nu_1-\nu_2)^2 (\nu_2-\nu_3)^2 (\nu_3-\nu_1)^2
\Bigr]^{1/2} 
\nonumber\\
&&  
=
\frac{8}{\omega \sin i (5+\cos^2 i)} 
Amp\left( h_{\rm Oct+C}^{+}|_{\omega} \right) ,  
\label{det-OctComega} 
\end{eqnarray}
where the R.H.S. is known up to this point. 

Equation ($\ref{hOct+Cplus}$) is rewritten as 
\begin{eqnarray}
r\times h_{\rm Oct+C}^{+} 
&=& 
-\frac{1}{4} m_{\rm tot}^2 \omega \sin i  
\nonumber\\
&&
\times 
\Bigl[
9 (1+\cos^2 i) 
\sqrt{27 \nu_1^2\nu_2^2\nu_3^2 
+ (\nu_1-\nu_2)^2 (\nu_2-\nu_3)^2 (\nu_3-\nu_1)^2} 
\cos (3\omega t - \Psi_{3\omega}) 
\nonumber\\
&&
{}\;\;\;
- 
\frac12
(5+\cos^2 i) 
\Bigl(
3 
\{ \nu_1 \nu_2 (\nu_1 - \nu_2) 
+\nu_2 \nu_3 (\nu_2 - \nu_3) 
+\nu_3 \nu_1 (\nu_3 - \nu_1) 
\}^2 
\nonumber\\
&& 
{}\;\;\;\;\;\; 
+ 
(\nu_1-\nu_2)^2 (\nu_2-\nu_3)^2 (\nu_3-\nu_1)^2
\Bigr)^{1/2}  
\cos(\omega t - \Psi_{\omega})
\Bigr] , 
\label{hOct+Cplus2}
\end{eqnarray}
where phases of $h_{\rm Oct+C}^{+}$ at $\omega$ and $3\omega$ 
are defined as 
\begin{eqnarray}
\tan\Psi_{\omega}
&=& 
-\frac{(\nu_2 - \nu_3) 
\{ (\nu_1 - \nu_2) (\nu_1 - \nu_3) - 3 \nu_2 \nu_3 \}}
{\sqrt3 \nu_1 
\{ \nu_2 (\nu_2 - \nu_1) + \nu_3 (\nu_3 - \nu_1) \} 
} , 
\label{det-Psiomega}
\end{eqnarray}
and 
\begin{eqnarray}
\tan\Psi_{3\omega}
&=& 
\frac{(\nu_1-\nu_2)(\nu_2-\nu_3)(\nu_3-\nu_1)}
{3^{3/2} \nu_1 \nu_2 \nu_3} , 
\label{det-Psi3omega}
\end{eqnarray}
respectively. 
These phases are not directly observable. 
Time lags between the mass quadrupole wave 
with frequency $2\omega$ 
and the combination of mass octupole and 
current quadrupole waves with $\omega$ (or $3\omega$) 
can be measured. 
They are defined as 
\begin{eqnarray}
\Delta t_{\omega} 
&\equiv& 
t_{\omega} - t_{\rm Q} 
\nonumber\\
&=&
\frac{2 \Psi_{\omega} - \Psi_{\rm Q}}{2\omega} , 
\label{Deltatomega}
\end{eqnarray}
\begin{eqnarray}
\Delta t_{3\omega} 
&\equiv& 
t_{3\omega} - t_{\rm Q} 
\nonumber\\
&=&
\frac{2 \Psi_{\omega} - 3 \Psi_{\rm Q}}{6\omega} , 
\label{Deltat3omega}
\end{eqnarray}
respectively, where $t_{\rm Q}$, $t_{\omega}$ and $t_{3\omega}$ 
are defined as 
\begin{eqnarray}
t_{\rm Q} &\equiv& \frac{\Psi_{\rm Q}}{2\omega} , 
\label{tQ}
\\
t_{\omega} &\equiv& \frac{\Psi_{\omega}}{\omega} , 
\label{tomega}
\\
t_{3\omega} &\equiv& \frac{\Psi_{3\omega}}{3\omega} .
\label{t3omega}
\end{eqnarray}

It is worthwhile to mention that $\Delta t_{\omega}$ 
and $\Delta t_{3\omega}$ are observable and thus gauge-invariant, 
whereas $t_{\rm Q}$, $t_{\omega}$ and $t_{3\omega}$ are gauge-dependent 
in a sense that they rely on a degree of freedom for 
choosing an initial time ``$t_0$''. 

In principle, time lags in the cross mode, phase of which 
is different from that of the plus mode by $45^{o}$, 
are the same as those in the plus mode, 
and thus bring no additional information 
on parameter determinations, 
though they may play a supplementary role in improving accuracy 
of a practical data analysis. 

As a result, one can determine five quantities 
$m_{\rm tot}$, $\nu_1$, $\nu_2$, $\nu_3$ and $r$ from 
Eqs. ($\ref{unity}$), ($\ref{mchirp}$), ($\ref{det-r2}$), 
($\ref{det-OctC3omega}$), ($\ref{det-OctComega}$), 
($\ref{Deltatomega}$) and ($\ref{Deltat3omega}$).  
In principle, five out of the seven equations are sufficient 
for parameter determinations if the source is known 
as the Lagrange's solution {\it a priori}. 
It should be noted that two remaining equations are {\it never} 
redundant but play an important role in checking 
whether a source is the particular three-body system or not. 
If the remaining equations are satisfied by the determined 
parameter values, one can safely say that 
the source is in Lagrange's orbit. 
If not, it could be other systems. 

\begin{figure}[t]
\includegraphics[width=14cm]{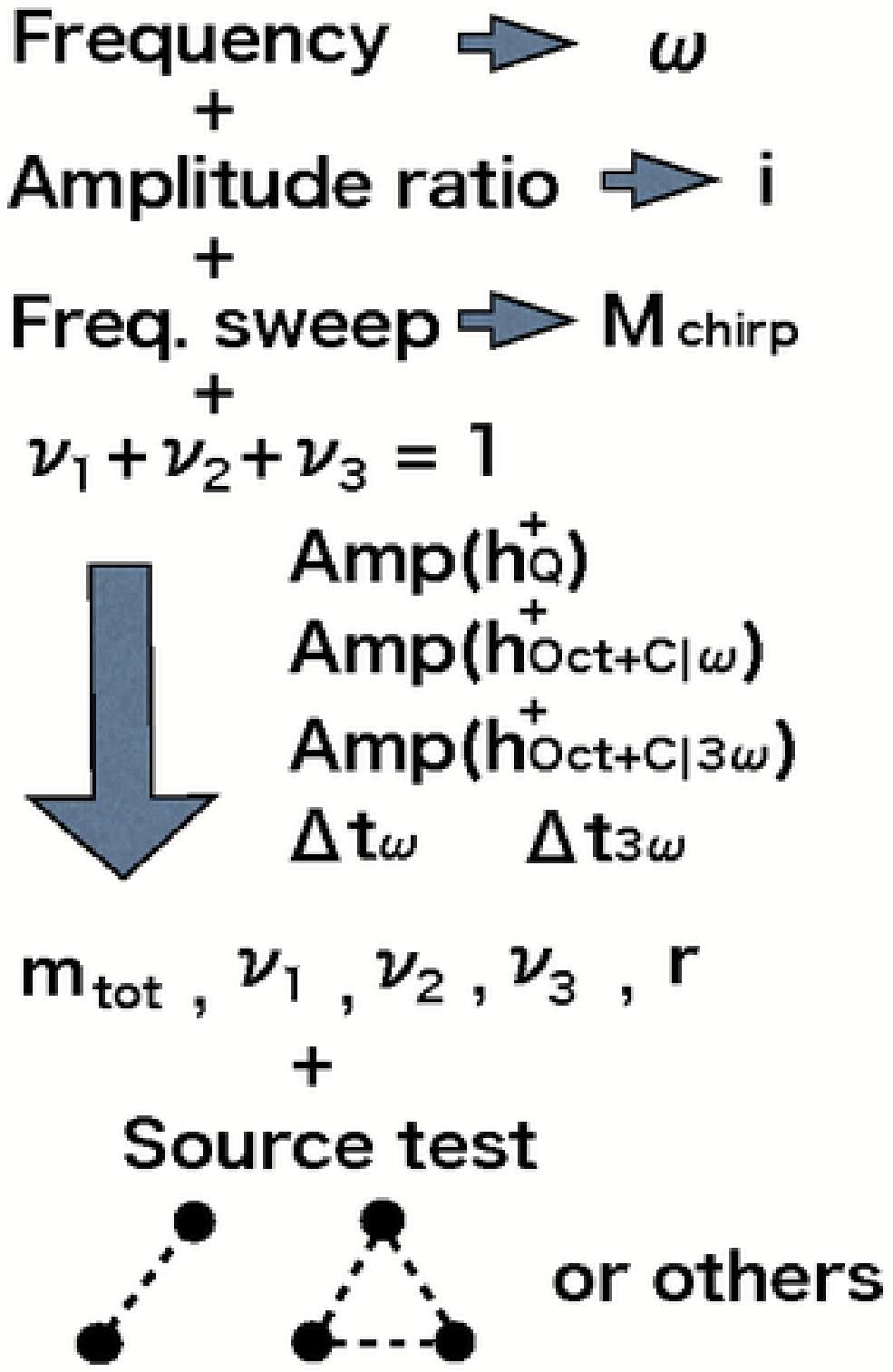}
\caption{
Flow chart of the parameter determinations 
and source tests.  
By using equations derived in this paper, 
the source parameters can be determined through 
gravitational-wave observations alone. 
In addition, a binary source can be distinguished from 
a three-body system in Lagrange's orbit or others. 
}
\label{f3}
\end{figure}

Figure $\ref{f3}$ shows a flow chart of the parameter determinations 
and possible source tests that are discussed above.

\begin{table}
\caption{
List of quantities characterizing a system in this paper. 
}
\label{table1}
  \begin{center}
    \begin{tabular}{ll}
\hline
Symbol & Definition \\
\hline 
$T$ & Orbital period \\
\hline 
$\omega$ & Angular velocity $(=2\pi/T)$ \\
\hline 
$a$ & Edge length of a Lagrange's equilateral triangle\\
\hline 
$a_p$ & Distance of each body from their center of mass\\
\hline 
$\theta_p$ & Initial angular position of each body\\
\hline 
$m_p$ & Mass of each body\\
\hline 
$\nu_P$ & Mass ratio of each body\\
\hline 
$m_{\rm tot}$ & Total mass\\
\hline 
$M_{\rm chirp}$ & Chirp mass\\
\hline 
$F$ & Ratio as $M_{\rm chirp} m_{\rm tot}^{-1}$\\
\hline 
$r$ & Distance of a source from an observer\\ 
\hline 
$h_{\rm Q}^{+, \times}$ & $+$ or $\times$ mode of 
mass quadrupolar waves\\
\hline 
$h_{\rm Oct}^{+, \times}$ & $+$ or $\times$ mode of 
mass octupolar waves\\
\hline 
$h_{\rm C}^{+, \times}$ & $+$ or $\times$ mode of 
current quadrupolar waves\\
\hline 
$h_{\rm Oct+C}^{+, \times}$ & $h_{\rm Oct}^{+, \times} + h_{\rm C}^{+, \times}$\\
\hline 
$Amp$ & Amplitude of a wave component\\
\hline
$\Psi_{\rm Q}$ & Phase of mass quadrupolar waves with $2\omega$\\
\hline
$\Psi_{\omega}$ & Phase of waves with $\omega$\\
\hline
$\Psi_{3\omega}$ & Phase of waves with $3\omega$\\
\hline
$t_{\rm Q}$ & Time corresponding to the phase $\Psi_{\rm Q}$\\
\hline
$t_{\omega}$ & Time corresponding to the phase $\Psi_{\omega}$\\
\hline
$t_{3\omega}$ & Time corresponding to the phase $\Psi_{3\omega}$\\
\hline
$\Delta t_{\omega}$ & Observable time lag defined as $t_{\omega}-t_{\rm Q}$\\
\hline
$\Delta t_{3\omega}$ & Observable time lag defined as $t_{3\omega}-t_{\rm Q}$\\
    \end{tabular}
  \end{center}
\end{table}

\section{Conclusion} 
We have considered the three-body wave forms 
at the mass quadrupole, octupole and current quadrupole orders, 
especially in an analytical method. 
By using the derived expressions, 
we have solved a gravitational-wave inverse problem of 
determining the source parameters 
to the particular configuration (three masses, 
a distance of the source to an observer, and 
the orbital inclination angle to the line of sight) 
through observations of the gravitational wave forms alone. 
We have discussed also whether and how a binary source 
can be distinguished from a three-body system 
in Lagrange's orbit or others, 
and thus proposed a binary source test. 

To be more precise, we should take account of post-Newtonian 
corrections to both wave generation and propagation   
beyond the linearized theory. 
It is interesting to consider different shapes of orbits and 
different numbers of bodies ($N=4$ or more). 
This is a topic of future study. 

\section*{Acknowledgment} 
We are grateful to P. Hogan for useful comments 
on the manuscript. 
We would like to thank E. Gourgoulhon  
for his useful conversations at Meudon observatory, 
where this work was partly done.  
This work was supported in part 
by a fellowship from Hirosaki University 
and by a Japanese Grant-in-Aid 
for Scientific Research from the Ministry of Education, 
No. 21540252. 

\appendix
\section{Lagrange's solution in complex variables}
\subsection{Correspondence between position angles and mass ratios}
Let each mass located at each vertex of 
a regular triangle with the side length denoted as $a$. 
For simplicity, the center of the complex plane $z \equiv x + i y$ 
is chosen tentatively as that of the triangle. 
Without loss of generality, we assume initial positions of 
three masses as $a$, $a e^{i 2\pi/3}$, $a e^{i 4\pi/3}$. 
The center of mass for the three masses is located at 
$z_{CM} = a 
\left( \nu_1 + \nu_2 e^{i 2\pi/3} + \nu_3 e^{i 4\pi/3} \right)$. 

When we consider gravitational waves, 
it is convenient to choose the coordinates origin as 
the mass center so that the dipole moment can vanish. 
In order to do so, we make a translation as 
$z \to z-z_{CM}$, 
for which the initial position of each mass becomes 
\begin{eqnarray}
z_{{\rm I} 1}&=& 
\frac{\sqrt{3}}{2} a
\left(\sqrt{3}(\nu_2+\nu_3) - (\nu_2-\nu_3)i\right) , 
\label{zI1}\\
z_{{\rm I} 2}&=& 
- \frac{\sqrt{3}}{2} a
\left(\sqrt{3} \nu_1 - (\nu_1+2\nu_3)i\right) ,  
\label{zI2}\\
z_{{\rm I} 3}&=& 
- \frac{\sqrt{3}}{2} a
\left(\sqrt{3} \nu_1 + (\nu_1+2\nu_2)i\right) ,  
\label{zI3}
\end{eqnarray}
where the subscript $I$ means the initial values. 
The magnitude of $z_{{\rm I} 1}$, $z_{{\rm I} 2}$ and $z_{{\rm I} 3}$, 
which are nothing but three masses' orbital radii 
around their center of mass, 
become 
\begin{eqnarray}
a_1 &\equiv& |z_{{\rm I} 1}| 
\nonumber\\ 
&=& a \sqrt{3 (\nu_2^2+\nu_3^2+\nu_2 \nu_3)} , 
\label{a1}\\
a_2 &\equiv& |z_{{\rm I} 2}| 
\nonumber\\
&=& a \sqrt{3 (\nu_3^2+\nu_1^2+\nu_3 \nu_1)} , 
\label{a2}\\
a_3 &\equiv& |z_{{\rm I} 3}| 
\nonumber\\
&=& a \sqrt{3 (\nu_1^2+\nu_2^2+\nu_1 \nu_2)} , 
\label{a3}
\end{eqnarray}
which respect symmetry for cyclic changes as 
$1 \to 2 \to 3 \to 1$. 

We obtain a relation between an position angle $\theta_1$ 
(with respect to the center of mass) and the mass ratios as 
\begin{eqnarray}
e^{i\theta_1} &=& 
\frac{z_{{\rm I} 1}}{|z_{{\rm I} 1}|} 
\nonumber\\
&=& \frac{\sqrt{3}(\nu_2+\nu_3) - (\nu_2-\nu_3)i}
{2 \sqrt{\nu_2^2+\nu_3^2+\nu_2 \nu_3}} , 
\label{eitheta1}
\end{eqnarray}
Similarly, we obtain 
\begin{eqnarray}
e^{i\theta_2} &=& 
- \frac{\sqrt{3} \nu_1 - (\nu_1+2\nu_3)i}
{2 \sqrt{\nu_3^2+\nu_1^2+\nu_3 \nu_1}} , 
\label{eitheta2} \\
e^{i\theta_3} &=& 
- \frac{\sqrt{3} \nu_1 + (\nu_1+2\nu_2)i}
{2 \sqrt{\nu_1^2+\nu_2^2+\nu_1 \nu_2}} . 
\label{eitheta3}
\end{eqnarray}

By using the complex representations,  
double-angle and triple-angle relations, 
which are useful for later computations,  
are obtained as 
\begin{eqnarray}
e^{2 i\theta_1} &=&
\frac{(\nu_2^2+\nu_3^2+4\nu_2 \nu_3) - \sqrt{3}(\nu_2^2-\nu_3^2)i}
{2 (\nu_2^2+\nu_3^2+\nu_2 \nu_3)} , 
\label{e2itheta1}\\
e^{2 i\theta_2} &=&
\frac{(\nu_1^2-2\nu_3^2-2\nu_3 \nu_1) - \sqrt{3}\nu_1(\nu_1+2\nu_3)i}
{2 (\nu_3^2+\nu_1^2+\nu_3 \nu_1)} , 
\label{e2itheta2}\\
e^{2 i\theta_3} &=&
\frac{(\nu_1^2-2\nu_2^2-2\nu_1 \nu_2) + \sqrt{3}\nu_1(\nu_1+2\nu_2)i}
{2 (\nu_1^2+\nu_2^2+\nu_1 \nu_2)} , 
\label{e2itheta3} 
\end{eqnarray}
and 
\begin{eqnarray}
e^{3 i\theta_1} &=&
\frac
{3^{3/2} \nu_2 \nu_3 (\nu_2+\nu_3) 
- (\nu_2-\nu_3)(\nu_2+2\nu_3)(2\nu_2+\nu_3)i}
{2 (\nu_2^2+\nu_3^2+\nu_2 \nu_3)^{3/2}} ,  
\label{e3itheta1}\\
e^{3 i\theta_2} &=&
\frac
{3^{3/2} \nu_3 \nu_1 (\nu_3+\nu_1) 
+ (\nu_1-\nu_3)(\nu_1+2\nu_3)(2\nu_1+\nu_3)i}
{2 (\nu_3^2+\nu_1^2+\nu_3 \nu_1)^{3/2}} ,  
\label{e3itheta2}\\
e^{3 i\theta_3} &=&
\frac
{3^{3/2} \nu_1 \nu_2 (\nu_1+\nu_2) 
- (\nu_1-\nu_2)(\nu_1+2\nu_2)(2\nu_1+\nu_2)i}
{2 (\nu_1^2+\nu_2^2+\nu_1 \nu_2)^{3/2}} ,  
\label{e3itheta3}
\end{eqnarray}
respectively.

Some useful relations are obtained as 
\begin{eqnarray}
A_{\rm C} &\equiv& 
\sum_p \nu_p |z_{{\rm I} p}|^3 \cos\theta_p 
\nonumber\\ 
&=& \frac{9}{2} \nu_1 
[ \nu_2^2 + \nu_3^2 - \nu_1 (\nu_2 + \nu_3) ] a^3 , 
\label{sumcos}\\
A_{\rm S} &\equiv& 
\sum_p \nu_p |z_{{\rm I} p}|^3 \sin\theta_p 
\nonumber\\
&=& \frac{3^{3/2}}{2} (\nu_2-\nu_3)  
[ \nu_1(\nu_1 - \nu_2 - \nu_3) - 2 \nu_2 \nu_3 ] a^3 ,   
\label{sumsin}
\end{eqnarray}
where we use Eqs. ($\ref{eitheta1}$)-($\ref{eitheta3}$). 
Therefore, we obtain 
\begin{equation}
\sum_p \nu_p |z_{{\rm I} p}|^3 \cos(\omega t + \theta_p) 
= \sqrt{A_{\rm C}^2 + A_{\rm S}^2} \cos(\omega t + \alpha) , 
\label{sumcosomegat}
\end{equation}
where we define 
\begin{equation}
\alpha \equiv \arctan\left(\frac{A_{\rm S}}{A_{\rm C}}\right) . 
\label{alpha}
\end{equation}

Next we consider $3 \omega$ parts. 
We obtain 
\begin{eqnarray}
B_{\rm C} &\equiv& 
\sum_p \nu_p |z_{{\rm I} p}|^3 \cos3\theta_p 
\nonumber\\ 
&=& 27 \nu_1 \nu_2 \nu_3 a^3 , 
\label{sumcos3}\\
B_{\rm S} &\equiv& 
\sum_p \nu_p |z_{{\rm I} p}|^3 \sin3\theta_p 
\nonumber\\
&=& - 3^{3/2} (\nu_1-\nu_2)(\nu_2-\nu_3)(\nu_3-\nu_1) a^3 ,   
\label{sumsin3}
\end{eqnarray}
where we use Eqs. ($\ref{e3itheta1}$)-($\ref{e3itheta3}$). 
Therefore, we obtain 
\begin{equation}
\sum_p \nu_p |z_{{\rm I} p}|^3 \cos3(\omega t + \theta_p) 
= \sqrt{B_{\rm C}^2 + B_{\rm S}^2} \cos(3\omega t + \beta) , 
\label{sumcos3omegat}
\end{equation}
where we define 
\begin{equation}
\beta \equiv \arctan\left(\frac{B_{\rm S}}{B_{\rm C}}\right) . 
\label{beta}
\end{equation}

\subsection{Chirp mass} 
By straightforward but lengthy calculations, 
one can show several identities as 
\begin{eqnarray}
\sum_p \nu_p \left( \frac{M^{\rm eff}_p}{m_{\rm tot}} \right)^{2/3} 
&=& 
\nu_1\nu_2 + \nu_2\nu_3 + \nu_3\nu_1 ,  
\end{eqnarray}
\begin{eqnarray}
\sum_{p \neq q} \nu_p \nu_q 
-\sum_p \nu_p \left( \frac{M^{\rm eff}_p}{m_{\rm tot}} \right)^{2/3} 
&=& \nu_1\nu_2 + \nu_2\nu_3 + \nu_3\nu_1 , 
\end{eqnarray}
\begin{eqnarray}
2 \sum_{p \neq q} \nu_p \nu_q 
\left( \frac{M^{\rm eff}_p}{m_{\rm tot}} \right)^{2/3}
\left( \frac{M^{\rm eff}_q}{m_{\rm tot}} \right)^{2/3}
\sin^2(\theta_p-\theta_q) 
&=& 3 \nu_1\nu_2\nu_3 ,  
\end{eqnarray}
where Eq. ($\ref{unity}$) is frequently used. 

By substituting these relations into Eq. ($\ref{Mchirp-old}$), 
we obtain 
\begin{eqnarray}
&&
\left( \sum_p \nu_p \left( \frac{M^{\rm eff}_p}{m_{\rm tot}} \right)^{2/3} \right)^2 
-2 \sum_{p \neq q} \nu_p \nu_q 
\left( \frac{M^{\rm eff}_p}{m_{\rm tot}} \right)^{2/3}
\left( \frac{M^{\rm eff}_q}{m_{\rm tot}} \right)^{2/3}
\sin^2(\theta_p-\theta_q) 
\nonumber\\
&&= 
\frac12 [\nu_1^2 (\nu_2-\nu_3)^2 + \nu_2^2 (\nu_3-\nu_1)^2 
+ \nu_3^2 (\nu_1-\nu_2)^2] . 
\end{eqnarray}

Therefore, we can prove that the chirp mass defined 
by Eq. ($\ref{Mchirp-old}$) is expressed 
in terms of only the mass ratios as Eq. ($\ref{mchirp}$).

\end{document}